\documentclass{iaus}
\usepackage{graphicx}
\usepackage{caption}

\title[] 
{Variability of Young Massive Stars in the Arches Cluster}

\author[]   
{K. Markakis$^1$$^,$$^2$
 \, A.Z. Bonanos$^1$$^,$$^2$
 \, G. Pietrzynski$^3$
 \, L. Macri$^4$
 \, K.Z. Stanek$^5$}

\affiliation{$^1$National Observatory of Athens, Institute of Astronomy \& Astrophysics, \\ I. Metaxa \& Vas. Pavlou St., P. Penteli 15236, Athens, Greece \\ {\tt markakis@astro.noa.gr}, {\tt bonanos@astro.noa.gr} \\[\affilskip]
$^2$K.M. \& A.Z.B. acknowledge support from the IAU and the European Commission \\for an FP7 Marie Curie International Reintegration Grant. \\[\affilskip]
$^3$Warsaw University Observatory, Al. Ujazdowskie 4, 00-478 Warszawa, Poland\\ Universidad de Concepci{\'o}n, Departamento de Astronomia,
Casilla 160-C, Concepci{\'o}n, Chile
\\[\affilskip]
$^4$Department of Physics \& Astronomy, Texas A\&M University, College Station, TX 77842-4242, USA\\[\affilskip]
$^5$The Ohio State University, 140 West 18th Avenue, Columbus, OH 43210, USA{}}

\pubyear{2010}
\volume{272}  
\pagerange{1--2}
\jname{Active OB stars: structure, evolution, mass loss\\ and critical limits} 
\editors{C. Neiner, G. Wade, G. Meynet \& G. Peters}
\begin{document}

\maketitle

\begin{abstract}
We present preliminary results of the first near-infrared variability study of the Arches cluster, using adaptive optics data from NIRI/Gemini and NACO/VLT. The goal is to discover eclipsing binaries in this young (2.5 $\pm$ 0.5 Myr), dense, massive cluster for which we will determine accurate fundamental parameters with subsequent spectroscopy. Given that the Arches cluster contains more than 200 Wolf-Rayet and O-type stars, it provides a rare opportunity to determine parameters for some of the most massive stars in the Galaxy.
\end{abstract}

\firstsection 
\section{Introduction}
One of the most important questions is how massive can the most massive stars in the Universe be today. In other words what is the upper limit of the Initial Mass Function in the Universe. The Arches Cluster provides us with a unique opportunity to address this question because it has all the criteria of the ideal place to look at for massive eclipsing binary systems. It lies near the Galactic Center which is a very dense region that benefits the formation of massive stars and the cluster itself is very young which can guarantee that its stars will not have evolved significantly. 
\vspace{-0.6cm}
\section{Datasets \& Reduction}
We used two datasets in the $K_s$ band. The first dataset was obtained with Gemini's NIRI infrared camera which consisted of 16 observations of 30 and 1 second exposures respectively covering 8 nights from April to July of 2006. The NIRI data are pending a linearity correction. The second dataset was obtained with the VLT's NACO infrared camera and consisted of 46 observations of 20 seconds exposures each covering 31 nights from June of 2008 to March of 2009. The reduction of the NIRI images was performed with the IRAF\footnote{IRAF is distributed by the NOAO, which are operated by the
Association of Universities for Research in Astronomy, Inc., under
cooperative agreement with the NSF.} Gemini v1.9 package while
the reduction of the NACO images was performed via the NACO reduction pipeline, based on ESO's Common Pipeline Library.
\firstsection
\section{Image Subtraction \& Photometry}
We used the image subtraction package ISIS \cite[(Alard \& Lupton 1998]{Alardlupton98}\cite[, Alard 2000)]{Alard00}, which is optimal for detecting variables in crowded fields, together with IRAF's DAOPHOT \cite[(Stetson 1987)]{Stetson87} package on the reference image from ISIS. Although ISIS allows for a spatially variable PSF it doesn't provide us with an accurate PSF model due to the anisoplanatic effects introduced by the imperfect correction of atmospheric turbulence by the adaptive optics. The light curve presented below has been obtained with ISIS and its overall shape strongly suggests that this may be a contact eclipsing binary. In order to obtain more accurate photometry and confirm our result, we are currently using the StarFinder code \cite[(Diolaiti et al. 2000a)]{Diolaiti00a}. StarFinder was designed to extract an empirical PSF from the image, that also takes into account the anisoplanatic effects caused by the adaptive optics.

\firstsection
\section{Preliminary results}
We present the light curve of an eclipsing binary candidate in the Arches cluster from the NACO data, which corresponds to the second star in the catalog of Figer et al. (2002). It has a 10.49 day period and is likely in a contact configuration. This candidate eclipsing binary has a spectral type of WN7 \cite[(Blum et al. 2001)]{Blumet_al01}. It has also been identified as radio source AR10 with a mass loss rate of the order of 1.9 x 10$^{-5}${\mbox{M$_{\odot}$}} yr$^{-1}$  \cite[(Lang et al. 2001b)]{Lang01b} and as X-Ray source A6, probably associated with the close pair of radio sources AR6 and AR10 \cite[(Wang et al. 2006)]{Wang06}. Finally it has an initial mass greater than 120 {\mbox{M$_{\odot}$}} \cite [(Figer et al. 2002)]{Figer2002}. 

	\begin{center}
	\includegraphics[scale=0.32]{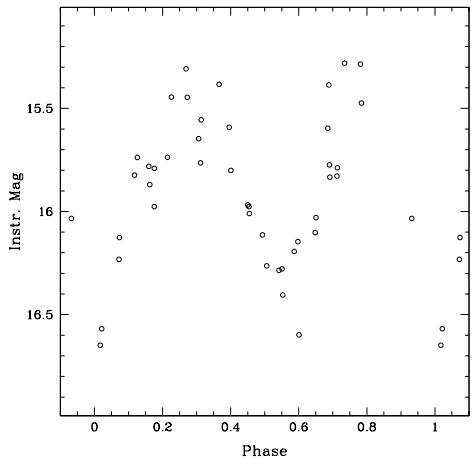}
	\captionof{figure}{$K_s-$band light curve of the candidate eclipsing binary in the Arches cluster.} 
	\end{center}

\end{document}